\documentclass[12pt]{article}

\usepackage[utf8]{inputenc} % allow utf-8 input
\usepackage[T1]{fontenc}    % use 8-bit T1 fonts
\usepackage{hyperref}       % hyperlinks
\usepackage{url}            % simple URL typesetting
\usepackage{booktabs}       % professional-quality tables
\usepackage{amsfonts}       % blackboard math symbols
\usepackage{nicefrac}       % compact symbols for 1/2, etc.
\usepackage{microtype}      % microtypography
\usepackage{lipsum}

\usepackage{ifpdf}
 \ifpdf
 \else
 \fi
\usepackage{amsmath}
\usepackage{algorithmic}
\usepackage{array}
\usepackage[numbers]{natbib}

\usepackage{stfloats}
\usepackage{amssymb}
\usepackage{mathdots}% http://ctan.org/pkg/mathdots

\usepackage{breqn}
\usepackage{amsfonts}
\usepackage{upgreek}
\usepackage{xcolor}

\usepackage{amsmath}
\usepackage{bm} % bold maths
\usepackage{amsthm}
\usepackage{amsfonts}
\usepackage{graphicx}
\usepackage{subfigure}
\usepackage{relsize}
\usepackage{subeqnarray}

\usepackage{arxiv}
\usepackage{upgreek}
\usepackage{epstopdf}
\newtheorem{definition}{Definition}

\begin{document}
\title{On-line parameter and state estimation of an air handling unit model: experimental results using the modulating function method }
%\author{Ana~Ionesi, Hossein~Ramezani
%        and~Jerome~Jouffroy
%        % <-this % stops a space
%\thanks{Ana Ionesi, Hossein Ramezani and Jerome Jouffroy are with SDU Mechatronics, the Mads Clausen Institute, University of Southern Denmark,
%        DK-6400 Alsion 2, S\o nderborg, Denmark
%        {\tt\small (e-mail: ai@mci.sdu.dk, ramezani@mci.sdu.dk,  jerome@mci.sdu.dk )}}%
%%\thanks{*This work is supported by Danfoss.}% <-this % stops a space
%%\thanks{Manuscript received April 19, 2005; revised August 26, 2015.}
%}
%
%

\author{
 Ana~Ionesi \thanks{ \texttt{ai@mci.sdu.dk}}, Hossein~Ramezani\thanks{\texttt{ramezani@mci.sdu.dk}}, Jerome Jouffroy \thanks{\texttt{jerome@mci.sdu.dk}}\\
  SDU Mechatronics,\\
 Mads Clausen Institute,\\
  University of Southern Denmark,\\
  DK-6400 Alsion 2, S\o nderborg, Denmark 
}

\maketitle

\begin{abstract}
This paper considers the on-line implementation of the modulating function method, for parameter and state estimation, for the model of an air-handling unit, central element of HVAC systems. After recalling the few elements of the method, more attention is paid on issues related to its on-line implementation, issues for which we use two different techniques. Experimental results are obtained after implementation of the algorithms on a heat flow experiment, and they are compared with conventional techniques (conventional tools from Matlab for parameter estimation, and a simple Luenberger observer for state estimation) for their validation.
\end{abstract}

%\begin{IEEEkeywords}
%IEEE, IEEEtran, journal, \LaTeX, paper, template.
%\end{IEEEkeywords}

% \fi
%\IEEEpeerreviewmaketitle

\section{Introduction}

The challenges of the clear evidence of climate change \cite{EDENHOFER201348} bring in the front-line new regulations which aim to reduce the green-house gas emissions in all sectors. A notable amount of emissions as well as energy consumption is accounted by buildings, both in the industrial and residential sector. According to \cite{REYHERNANDEZ201885} the need for heating is going to decrease while the cooling demand in  buildings is going to increase in the coming decades due to the climate change. In this respect, heating ventilation and air conditioning  (HVAC)  systems with higher energy efficiency and better building designs are continuously researched and developed. 

To facilitate the transition towards reducing the energy consumption and the green-house gas emissions, mathematical models of buildings have been intensively used. The simulation tools available today \cite{CRAWLEY2008661} and the ongoing research offer a high variety of possibilities when it comes to design optimization \cite{KHEIRI2018897}, renewable integration \cite{ZHOU2018430}, retrofitting \cite{FAN20182140}, on-line fault detection diagnosis \cite{BONVINI2014156} and even optimized real-time control.

The HVAC system is one very important part of the building, which in Europe is estimated to share $76\%$ of the total energy use \cite{gruber2014alternative}. The duty of this system is to ensure a comfortable indoor environment by regulating the temperature and the air quality. The well-functioning of the HVAC system will contribute also to an optimal energy consumption. Several subsystems are considered when modeling an HVAC system, as explained in \cite{satyavada2016integrated}. Among them, the air handling unit (AHU) is the subsystem whose role is to condition the air circulated through the building. Different models have been proposed in the literature for modeling an AHU: physical models described by partial differential equations used by most of the simulation tools available \cite{CRAWLEY2008661}, transfer functions \cite{AFROZ2017}, data-driven models \cite{afram2014review}, and hybrid models or reduced-order models (ROM) \cite{afram2015gray}.

The latter lead to many possibilities as they easily link to estimation and control scenarios where real-time capabilities are important. In this paper, we investigate the use of the so-called modulating function method to perform both state and parameter estimation on a continuous-time ROM based on a simple thermal-electrical analogy \cite{fraisse2002development}. In comparison with more conventional approaches using discrete-time such as Kalman filtering, performing both state and parameter estimation in continuous-time relates better to the formalism in which the system is originally modeled and ensures the convergence of the estimates to the real values when sampling time approaches zero \cite{unbehauen1997identification}. Compared to their discrete-time counterparts estimating either parameters or state components is the necessity of considering time derivatives. The modulating function technique, originally introduced by Shinbrot \cite{shinbrot1957analysis}, allows to circumvent this issue elegantly, with the use of fixed-length time integrals of the measured signals, akin continuous-time FIR filtering. This offers the additional advantages of giving estimates after a fixed and predetermined amount of time, contrary to usual Kalman-based filtering methods. Originally proposed towards parameter estimation, the modulating function method was more recently extended to include state estimation \cite{liu2014non} \cite{jouffroy2015finite}.

Many studies considering the modulating function approach showed satisfactory performance, but mostly in simulations using artificial data \cite{jouffroy2015finite} or sources generated from real profiles \cite{asiri2017moving}. In contrast to that and to the best of our knowledge, not many applications can be found in the literature dealing with experiments done on actual measurements, to the notable exception of \cite{daniel1998experimental}, where Hartley modulating functions are used to estimate the parameters of a thyristor driven DC-motor. 

%The main concept of the modulating function is to multiply the measured signals with a well-known function which has specific characteristics \cite{Szabo1983}. In this way it is possible to transform the expression defined by the input-output signals, into a linear system of equations. Unlike other popular estimation methods as for example Kalman Filter, this method is very easy to understand because of its simplicity in terms of formulation and implementation. Moreover, not needing the initial conditions gives a great advantage over more conventional approaches and makes it an interesting alternative. By looking at the method from a receding horizon perspective, as we do in the present paper, facilitates the usage of the data once it is available. This results in having an on-line estimation scheme by shifting the integration window one step further each iteration. 

This paper addresses the question of real-time deterministic parameter and state estimation of an air handling unit using the modulating function method. The corresponding algorithms are implemented in Matlab/Simulink on an actual heat flow model and experimental results based on noisy measurements are presented. While the model itself is easy to develop and implement, performing estimation of its parameters and states is still a challenge given the actual distributed nature of these parameters, which are also subject to possible changes over time. In doing so, we also look at practical aspects related to implementation issues for using the modulating function method in an on-line situation, and which we believe has been less considered in the literature. Preliminary results were reported in \cite{AIM_ANA_Jerome}.

%Therefor, the contribution of the paper is to experimentally apply the modulating function method for parameter and state estimation on a ROM of an AHU that is performed on-line, directly on noisy measurements. For validation purpose the method is implemented and tested on the Quanser Heat Flow Experiment.

After this introduction, we start this paper by briefly explaining the principle of an HVAC system and that of an AHU, and give a simple reduced-order model (ROM) of the latter (section \ref{section:AHU}). Then, the basics of the modulating function method are recalled for both parameter and the state estimation in a state-space context (section \ref{section:MF}). A following section is more specifically addressed to issues that are more prone to arise in an on-line scenario, i.e. possible singularities coming from the choice of the modulating function, and ease of implementation of the moving-horizon version in a block diagram environment such as Simulink. For this, we use two different techniques, each one dedicated to these two specific issues (section \ref{section:choosingMF}). We validate the proposed techniques on a heat flow model and discuss our experimental results (section \ref{section:casestudy}). Brief concluding remarks end this paper (section \ref{section:conclusions}).

\begin{figure}[!htb]
	\centerline{
		\includegraphics[ width=8.8cm]{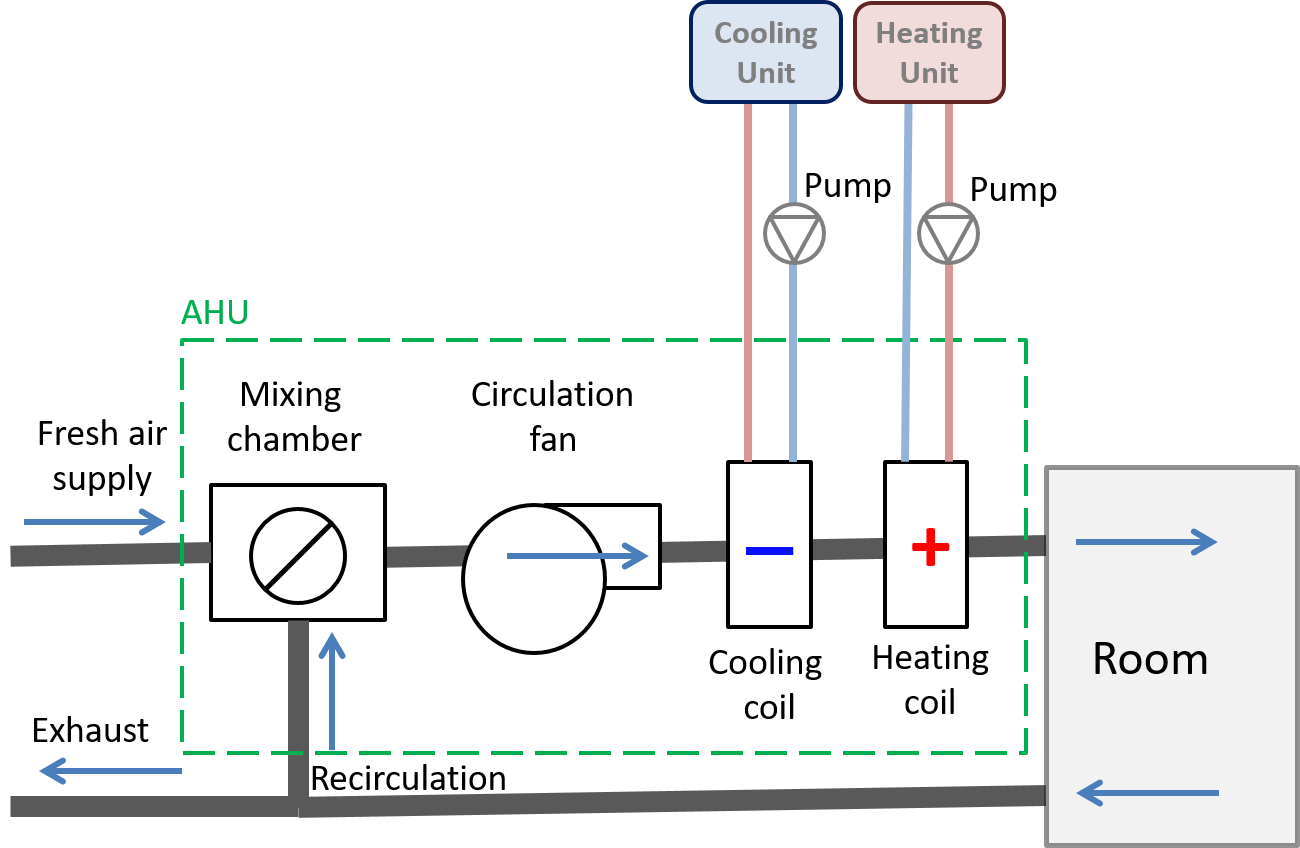}
	}
	\caption{Air flow diagram through a typical fan coil compact HVAC}
	\label{hvac}
\end{figure}

\section{Heat Flow Model of an AHU}
\label{section:AHU}

%In this section we will present the considered heat flow model of an AHU. First, we describe the components of the system and afterwards we give the details of the ROM mathematical expression. 

\subsection{System description}
\label{section:system}
The HVAC system within a building can be configured in many different ways with regards to the buildings' layout and purpose, application or users' needs. It usually contains a variety of components such as chillers, cooling towers, condensing boilers, AHUs, heat pumps, heat recovery units, fans, control valves, pipes, etc. as described, for example, in \cite{satyavada2016integrated}. The air flow in a typical compact HVAC system can be represented by the schematic diagram shown in Fig. \ref{hvac}. The heating and cooling coils are part of separate networks with complex units that deliver cold or heat based on requirements. Similar representations can be seen in \cite{DEY2016177} or \cite{chen2014development}.

An important component of the HVAC system is the AHU (marked with green dashed box in Fig. \ref{hvac}). The role of this component is to condition the air that is used for the ventilation of the rooms in order to maintain a comfortable indoor environment in terms of temperature and air quality. In this unit, the fresh air is mixed with air coming from the rooms in a mixing chamber, then is circulated by a constant air volume flow fan over a cooling/heating coil. Based on the needs, the air is heated or cooled accordingly, so that the temperature at the exit of the unit achieves prescribed values.

\subsection{Mathematical modeling}

Many models have been developed to represent the dynamic heat flow behavior of an AHU. In the existing literature, the models that have been developed range from the use basic physics \cite{setayesh2015comparison} to advanced data mining algorithms such as artificial neural networks and other machine learning techniques \cite{afram2018development}. An overview of the variety of these models can be found in \cite{AFROZ2017}.  

\begin{figure}[!htb]
	\centerline{
		\includegraphics[trim = 0mm 5mm 3mm 3mm, clip, width=8cm]{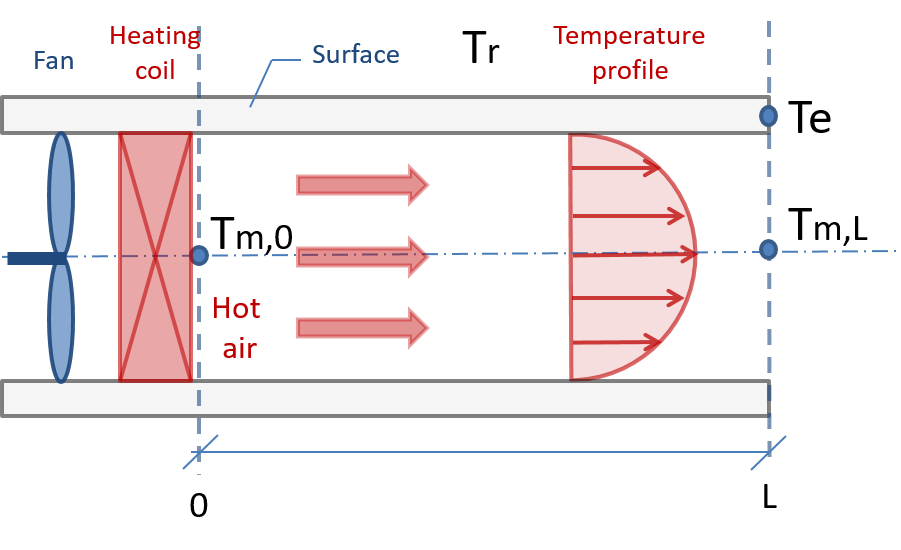}
	}
	\caption{Internal heat flow profile}
	\label{duct}
\end{figure}

The heat flow which is considered here is schematically represented in Fig. \ref{duct}. It simply consist of a fan, a heating coil, and a duct where temperature sensors can be placed. As expected from basic considerations, the temperature profile of the flowing medium, as well as the heat transfer coefficients typically varies with the length $L$ of the duct (see for example \cite{bergman2011fundamentals}). However, in the present application, only the temperature at the exit of the unit is of interest as it is typically the one which one wants to regulate. Hence we will, in the following, consider the measured temperature $T_{m,L}$, hereafter simply shortened as $T_{m}$. In order to model this heat flow, we resort to second-order linear ROM. A similar albeit first-order model is described in \cite{afram2015gray}.

\begin{figure}[!htb]
\centerline{
    \includegraphics[trim = 1mm 1mm 1mm 1mm, clip, width=7cm]{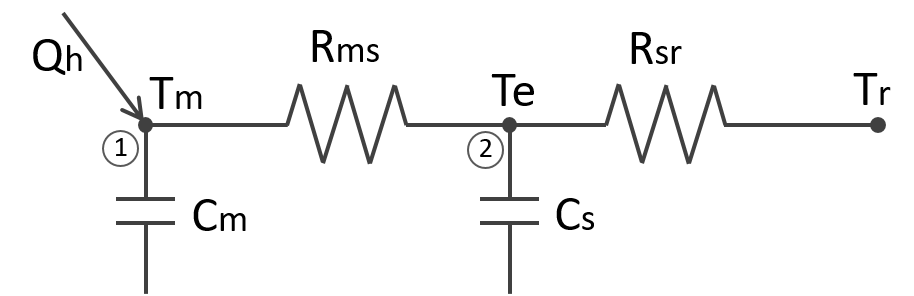}
    }
\caption{RC network analogy of the heat flow}
\label{RCNet}
\end{figure}

The proposed 2-nd order linear model uses the resistance capacitor (RC) network analogy depicted in Fig. \ref{RCNet}. In order to develop the model, we first write a flow balance equation for each node. Hence, we get  
\begin{equation}
C_{m}\frac{dT_{m}}{dt} =\frac{1}{R_{ms}}(T_{e}-T_{m})+Q_{h}
\label{eq:1}
\end{equation}
for node 1, representing the air inside the unit, where $T_{m}$ is the measured temperature in the AHU and $T_{e}$ is the envelope/surface temperature of the AHU, while $Q_{h}$ is the heating/cooling load added in the AHU. The constant parameters of this flow balance equation are $C_{m}$, the summed heat capacity of the sensor and of the air inside the unit and $R_{ms}$, the thermal resistance between the temperature sensor and the envelope of the unit. \\
For node 2, the surface/envelope of the unit, we have the differential relation
\begin{equation}
C_{s}\frac{dT_{e}}{dt} = \frac{1}{R_{ms}}(T_{m}-T_{e}) + \frac{1}{R_{sr}}(T_{r}-T_{e}) \\
\label{eq:2}
\end{equation}
where $T_{r}$ is the temperature of the room surrounding the unit. In equation (\ref{eq:2}), constant parameters are $C_{s}$, the heat capacity of the envelope of the AHU, $R_{ms}$, the thermal resistance between the sensor and the surface of the unit, and $R_{sr}$, the total thermal resistance between the surface of the unit and the room.

From (\ref{eq:1})-(\ref{eq:2}), it is simple to obtain the state-space representation
\begin{align}
\dot {\mathbf{x}} & = \mathbf{Ax}+ \mathbf{Bu} \label{eq:3} \\ 
y & =\mathbf{Cx} \label{eq:measure}
\end{align}
where, defining the state vector and the input vector as
\begin{equation}
\mathbf x = \left[ T_{m},T_{e} \right]^T,
\end{equation}
and
\begin{equation}
\mathbf u = \left[ Q_h , T_r \right]^T
\end{equation}
while assuming we only measure the temperature at the end of the unit, i.e. $y=T_m$, we have the matrices
\begin{equation}
\mathbf A = 
\begin{bmatrix} 
-\frac{1}{C_{m}R_{ms}}&\frac{1}{C_{m}R_{ms}}  \\ 
 \frac{1}{C_{s}R_{ms}}&-(\frac{1}{C_{s}R_{ms}}+\frac{1}{C_{s}R_{sr}}) \\ 
\end{bmatrix},
\end{equation}
\begin{equation}
\mathbf B = 
\begin{bmatrix} 
\frac{1}{C_{m}}& 0\\ 
0& \frac{1}{C_s R_{sr}}\\ 
\end{bmatrix},
\end{equation}
\hspace{2mm}
\begin{equation}
\mathbf C = 
\begin{bmatrix} 
1& 0 
\end{bmatrix}.
\label{Cmatrix}
\end{equation}

\section{The Modulating Function Method: parameter and state estimation}
\label{section:MF}

In this section, we briefly recall the basics of the modulating function method. For the sake of clarity, we do so on single-input single-output systems, with the multiple-input case as a direct extension.

%The core theory of the method can be tracked back to \cite{shinbrot1957analysis}, while recent applications of the modulating function can be found in  \cite{gavrilenko2017identification}, \cite{jouffroy2015finite} or \cite{noack2016fixed}.

The modulating function method being primarily based on an input-output description of a dynamical system, we start, as a preliminary by transforming our state-space presentation (\ref{eq:3})-(\ref{eq:measure}). Hence, considering the $n$-dimensional case where $\mathbf{x} \in \mathbb{R}^n$, $u \in \mathbb{R}$ and $y \in \mathbb{R}$, we differentiate the output equation \eqref{eq:measure} $n-1$ times and get the well-known expression
\begin{equation}
\bar{\mathbf{y}}= \mathbb{O} \mathbf{x}+ \mathbf{T} \bar{\mathbf{u}}
\label{eq:compactE} 
\end{equation}
where $ \bar{\mathbf{y}}=[y, \dot{y}, ... , y^{(n-1)}]^T $ and $ \bar{\mathbf{u}}=[u, \dot{u}, ... , u^{(n-1)}]^T $. Matrix $\mathbb{O}$ is the well-known observability matrix of the Kalman criterion and matrix $\mathbf{T}$ is the system specific Toeplitz matrix given by
\begin{equation}
\mathbf T=
\begin{bmatrix} 
\mathbf 0  & \mathbf  0& \hdots& \mathbf  0\\
\mathbf C \mathbf B  &\mathbf 0& \ddots& \vdots\\
\vdots  & \vdots&\ddots& \mathbf 0\\
\mathbf C \mathbf A^{n-2} \mathbf B  & \mathbf C \mathbf A^{n-3} \mathbf B&\dots& \mathbf  0\\
\end{bmatrix}.
\end{equation}

%The observability of the system can be verified by computing the determinant, $det(\mathbb{O})$, which must be different from 0. This is necessary in order to ensure that $\mathbb{O}^{-1}$ exists.

Then, differentiating equation \eqref{eq:measure} one more time and isolating $\mathbf{x}$ in expression \eqref{eq:compactE}, we get
\begin{equation}
y^{(n)}= \mathbf{C} \mathbf{A}^{n} \mathbb{O}^{-1}(\bar{\mathbf{y}}- \mathbf{T} \bar{\mathbf{u}}) +  \mathbf C \mathbb{C}^R \bar{\mathbf{u}} \\
\label{eq:ssrep} 
\end{equation}
where the reversed controllability matrix $\mathbb{C}^R$ is given by
\begin{equation}
\mathbb{C}^R=[\mathbf A^{(n-1)} \mathbf B, \mathbf A^{(n-2)} \mathbf B, ... ,\mathbf A \mathbf B,  \mathbf B]. 
\end{equation}
Note that an obvious condition for expression (\ref{eq:ssrep}) to be defined is the observability of state-space representation (\ref{eq:3})-(\ref{eq:measure}).
We thus have the input-output form
\begin{equation}
	y^{(n)}= -\mathbf{a}^T \bar{\mathbf{y}} +\mathbf{b}^T \bar{\mathbf{u}}
	\label{inputoutput}
\end{equation}
where $\mathbf{a}^T = \mathbf C \mathbf A^{n}\mathbb{O}^{-1 } = [a_0, a_1, ..., a_{n-1}]$ and $\mathbf{b}^T = \mathbf C \mathbb{C}^R - \mathbf C \mathbf A^{n}\mathbb{O}^{-1 } \mathbf T = [b_0, b_1, ..., b_{n-1}]$. Defining then the vector $\mathbf{Y}$ as $\mathbf{Y}= [\bar{\mathbf{y}}^T, \bar{\mathbf{u}}^T]^T$, the following parametric form is obtained
\begin{equation}
y^{(n)}= \mathbf Y^T \boldsymbol{\uptheta},
\label{eq:system}
\end{equation}
where the unknown parameter vector $\boldsymbol{\uptheta} \in \mathbb{R}^{2n}$ is given by $\boldsymbol{\uptheta}= [-\mathbf a^T , \mathbf b^T]^T$.
In case a constant and unknown disturbance $d$ is impacting the system's behavior at the same level as the input, expression (\ref{eq:system}) can be simply modified such that $\mathbf{Y}_d=[1,\mathbf{Y}^T]^T$ replaces $\mathbf{Y}$ and $\boldsymbol{\uptheta}_d = [d,\boldsymbol{\uptheta}^T]^T$ replaces $\boldsymbol{\uptheta}$.

Before proceeding, let us first recall the definition of a modulating function (see \cite{liu2014non}, \cite{jouffroy2015finite}).

\begin{definition}
	\label{MFdefinition}
The sufficiently smooth function $\varphi :  [0,T]\rightarrow \mathbb{R}$
is called a modulating function (of order $k$) if at least one of its boundaries and its derivatives up to order $k$ equals zero, i.e. if 
\begin{equation}
\varphi ^{(i)} (0)\cdot \varphi ^{(i)} (T) = 0 , i=\overline{0,k-1}.
\end{equation}
A modulating function where $\varphi ^{(i)} (0) = 0$ and $\varphi ^{(i)} (T) \neq 0$ (for $i=\overline{0,k-1}$) is called a left modulating function, while a modulating function with $\varphi ^{(i)} (0) \neq 0$  and $\varphi ^{(i)} (T) =0$  is called a right modulating function. If a modulating function is such that we have $\varphi ^{(i)} (0) = \varphi ^{(i)} (T) = 0$, then it is called a total modulating function.  
\end{definition}

Note a more general definition of a modulating function (see \cite{jouffroy2015finite}) allows to include expanding horizons, allowing thus to include alternative integral approaches \cite{mathew1972identification} \cite{saha1980identification}.

In this paper, we are interested in on-line estimation. Regarding parameter estimation, we hereby use the following receding-horizon integral operator on the signal $y(t)$, using a total modulating function given by \cite{noack2018road}
\begin{equation}
	L^i[y] := \int_{t-T}^{t} (-1)^i \varphi^{(i)}(\tau - t + T) y(\tau) d\tau.
	\label{Lintegraloperator}
\end{equation}
where $i=\overline{0,k-1}$ and $T>0$. Due to the fact that $\varphi(t)$ is a total modulating function and by simple integration by parts, operator (\ref{Lintegraloperator}) has the important property that 
\begin{equation}
	L^0[y^{(i)}] = L^i[y].
	\label{Loperatorproperty}
\end{equation}
Then, applying operator $L^0[\cdot]$ on each time-varying signal of (\ref{eq:system}), and using property (\ref{Loperatorproperty}), we are able to avoid both explicit time-derivatives of measured signals and unknown initial conditions to get
\begin{equation}
z = \mathbf w^T \boldsymbol{\uptheta},
\label{eq:systemOfE}
\end{equation}
where
\begin{equation}
 z= L^n[y]
\label{eq:z}
\end{equation}
and
\begin{equation}
	\mathbf w = [L^0[y],L^1[y],...,L^{n-1}[y],L^0[u],L^1[u],...,L^{n-1}[u]]^T.
	\label{eq:wi}
\end{equation}
Finally, we can obtain an estimate of parameter vector $\boldsymbol{\uptheta}$ by either proceeding to a conventional receding-horizon Gramian-based estimator \cite{5400719} \cite{reger2008algebraische} over a horizon $T'>0$, or aggregate a sufficient number of expressions (\ref{eq:systemOfE}), each one obtained with a different total modulating function, so that with $m_t \geq 2n$ total modulating functions, we get the set of linear equations
\begin{equation}
 \mathbf z = \mathbf W^T \boldsymbol{\uptheta}
\label{eq:est}
\end{equation}
where $ \mathbf z = [ z_1, z_2, ... ,  z_{m_t}]^T $ and $ \mathbf W = [\mathbf w_1, \mathbf w_2, ..., \mathbf w_{m_t}]$. In this case, we can in principle directly get the parameter vector estimate by computing
\begin{equation}
\hat {\boldsymbol{\uptheta}} = \left( \mathbf W \mathbf W ^T \right)^{-1}\mathbf W \mathbf z
\label{eq:leastsq}
\end{equation}
or, alternatively, proceed by adding another receding-horizon stage similar to the Gramian-based estimator alluded to above in order to remove noise further, and define the estimate as
\begin{equation}
	\hat {\boldsymbol{\uptheta}} = \left(\int_{t-T'}^{t} \mathbf W(\tau) \mathbf W^T(\tau) d\tau  \right)^{-1}\int_{t-T'}^{t} \mathbf W(\tau) \mathbf z(\tau) d\tau.
	\label{Gramianest}
\end{equation}

Turning now to state estimation, we resort to an integral operator based, this time, on a left modulating function, and given by
\begin{equation}
L^i_l[y] := \int_{t-T}^{t} (-1)^i \varphi_l^{(i)}(\tau - t + T) y(\tau) d\tau,
\label{Llintegraloperator}
\end{equation}
where $\varphi_l(t)\in [0,T]$ is a left modulating function. Because of the fact that $\varphi_l(T) \neq 0$, we have
\begin{equation}
	L^0_l[y^{(i)}] = \sum_{k=0}^{i-1} (-1)^k \varphi_l^{(k)}(T)y^{(i-1-k)}(t) + L^i_l[y].
\end{equation}
Proceeding then by applying operator $L^0_l[.]$ on each signal of (\ref{inputoutput}), we get the expression
\begin{equation}
	\boldsymbol{\upvarphi}_l \bar{\mathbf{y}} + \mathbf{a}^T \boldsymbol{\Gamma}_l \bar{\mathbf{y}} - \mathbf{b}^T \boldsymbol{\Gamma}_l \bar{\mathbf{u}} = 
	\mathbf{b}^T \mathbf{L}_l[u] -\mathbf{a}^T \mathbf{L}_l[y]-L_l^n[y]
	\label{xequation}
\end{equation}
where 
\begin{equation}
\boldsymbol{\upvarphi}_l =[(-1)^{n-1} \varphi_l^{(n-1)}(T),(-1)^{n-2} \varphi_l^{(n-2)}(T), \hdots, (-1)^{0} \varphi_l^{(0)}(T)],
\end{equation}
\begin{align}
\boldsymbol{\Gamma}_l\!=\!
\arraycolsep3.5pt\mbox{\small$ 
	\begin{bmatrix}
	0        & 0  & \cdots & 0 \\[0.4ex]
	(-1)^0 \varphi_l^{(0)}(T)       & 0  & \ddots & \vdots \\[0.4ex]
	\vdots & \ddots & \ddots & 0 \\[0.4ex]
	(-1)^{n-2} \varphi_l^{(0)}(T)                   & \hdots & (-1)^0 \varphi_l^{(0)}(T) & 0   
	\end{bmatrix}$}
\end{align}
and
\begin{equation}
\mathbf L_l[y] = \left[ L^0_l[y],L^1_l[y],...,L^{n-1}_l[y] \right]^T
\label{L_lvector}
\end{equation}
(and similarly for $\mathbf L_l[u]$).\\
Noticing then that 
\begin{equation}
	\left( \boldsymbol{\upvarphi}_l +\mathbf a^T \boldsymbol{\Gamma}_l \right) \mathbf T = \mathbf b^T \boldsymbol{\Gamma}_l,
\end{equation}
expression (\ref{xequation}) can be put into a form similar to (\ref{eq:systemOfE}), i.e. we have
\begin{equation}
z_l =  \mathbf w_l^T \left( \bar{\mathbf{y}} - \mathbf T \bar{\mathbf{u}} \right),
\label{xequationmodified}
\end{equation}
where
\begin{equation}
z_l = \mathbf{b}^T \mathbf{L}_l[u] -\mathbf{a}^T \mathbf{L}_l[y]-L_l^n[y]
\label{xequati}
\end{equation}
and
\begin{equation}
	\mathbf w_l = \left( \boldsymbol{\upvarphi}_l +\mathbf a^T \boldsymbol{\Gamma}_l \right)^T.
\end{equation}
Combining then $m_l \geq n$ left modulating functions, we obtain, similarly to (\ref{eq:est}), the expression
\begin{equation}
	\mathbf z_l =  \mathbf W_l^T \left( \bar{\mathbf{y}} - \mathbf T \bar{\mathbf{u}} \right),
	\label{xequationmodified2}
\end{equation}
which, defining the state corresponding to an observability canonical form as
\begin{equation}
	\bar{\mathbf x} = \bar{\mathbf{y}} - \mathbf T \bar{\mathbf{u}},
\end{equation}
leads to its estimate $\hat{\bar{\mathbf{x}}}$ with an expression similar to (\ref{eq:leastsq}). Alternatively, we can also estimate the original state of the system (\ref{eq:3})-(\ref{eq:measure}) by either using (\ref{eq:compactE}) or the simple transformation
\begin{equation}
\hat{\mathbf{x}} = \mathbb{O}^{-1} \hat{\bar{\mathbf x}}.
\label{xhateq}
\end{equation}

\section{Implementing the modulating function method for on-line applications}
\label{section:choosingMF}

\subsection{Using time-varying modulating functions}
\label{MFcalculation}

\begin{figure*}[htb!]
	\centering
	\includegraphics[width=0.98\columnwidth]{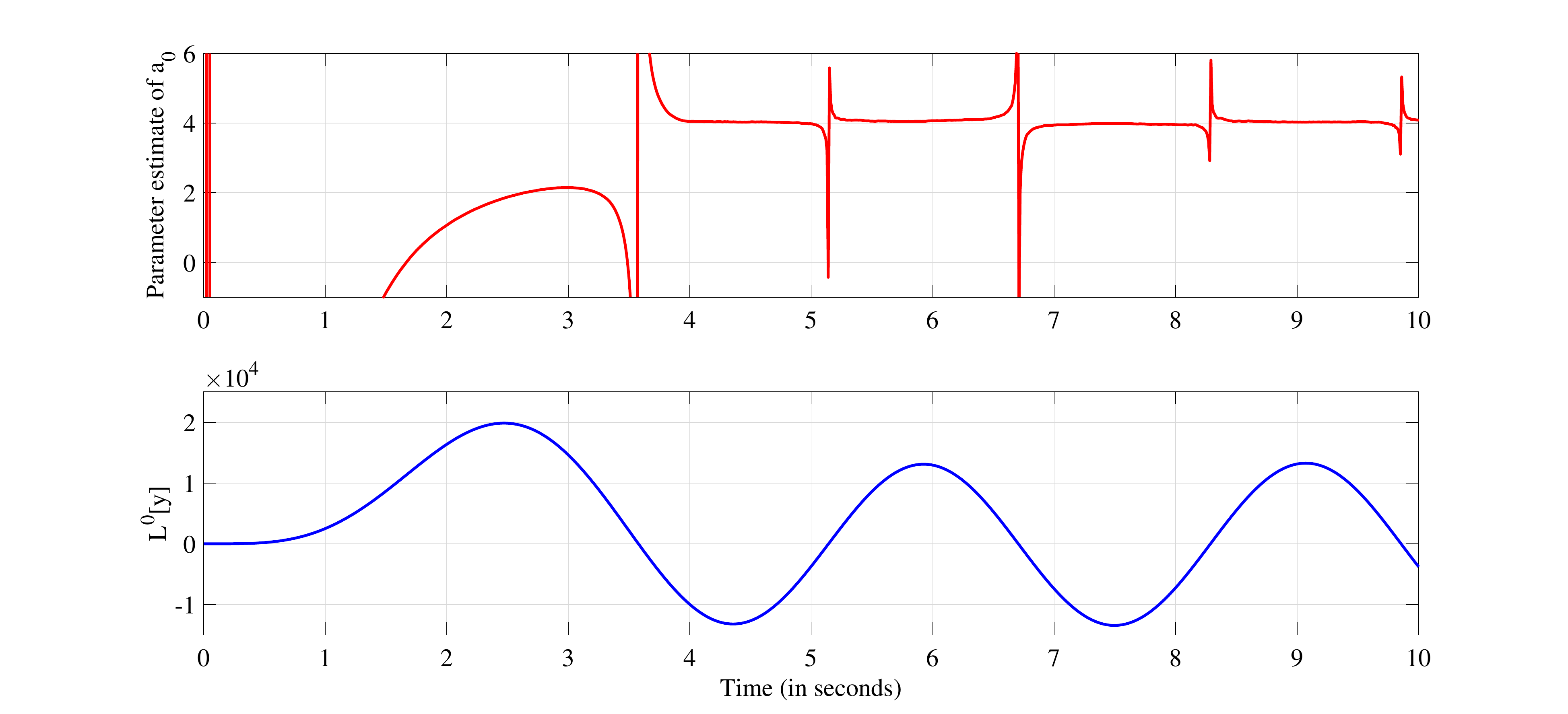}
	\caption{Parameter estimate $\hat{a}_0$ (top) and operator $L^0[y]$ (bottom) for the oscillator example (\ref{sinewave}).}
	\label{MFsinewave-fig}
\end{figure*}

Early works on the use of the modulating function method for offline identification of unknown parameters (see for example \cite{pearson1985identification}) start with the definition of a set of specific and predefined modulating functions, which can take different forms (ie Hartly, splines, trigonometric functions, etc). Once a sufficient number of modulating functions are used, one can get a parameter estimate directly by simple inversion, as in (\ref{eq:est}). However, when considering an on-line and receding-horizon situation, this can generally lead, especially with the presence of noise on the measurements, to singularity issues. For example, considering the trivial system
\begin{equation}
	\ddot{y} + a_0 y = 0,
	\label{sinewave}
\end{equation}
applying the modulating function method would simply lead to
\begin{equation}
	\hat{a}_0 = -\frac{L^2[y]}{L^0[y]},
	\label{a0estimate}
\end{equation}
where the denominator of (\ref{a0estimate}) could create difficulties. As an illustration, we have simulated the sinusoidal signal $y(t)=15 \sin2t$, corrupted with a uniform noise. As can be seen in Fig. \ref{MFsinewave-fig}, the estimate of $a_0$, obtained with the fixed total modulating function $\varphi(t)=t^2(t-T)^2$, shows singularities around the time when the signal $L^0[y](t)$ has its zero-crossings (see also similar spiking behaviors in Figure 3 of \cite{co1997batch}). Seeing $L^0[y](t)=<\varphi,y>$ as a dot product between two functions, one has therefore an orthogonality issue between these functions.

While one way to go around this issue typically consists of using a gradient or recursive least-squares algorithm (or even using more modulating functions), another possibility is to continuously redefine the modulating functions based on the current batch of the measured signals $y(t)$ and $u(t)$.\\
Indeed, setting the number of total modulating functions to $m_t=2n$, we impose a normalizing constraint represented as
\begin{equation}
\mathbf W = \mathbf I_{m_t}\text{.}
\label{identityconstraint}
\end{equation}
In this way, we are simply bypassing the matrix inversion of (\ref{eq:est}) or (\ref{Gramianest}). Realizing this constraint hence consists of finding the set of $m_t$ modulating functions that will fulfil (\ref{identityconstraint}). Since $\mathbf{W}$ is composed of $m_t$ vectors (\ref{eq:wi}), where $L^i[y]$ is given by (\ref{Lintegraloperator}), we have a set of $m_t$ integro-differential equations, where the unknowns are the $m_t$ total modulating functions $\varphi_k$'s, which are also constrained at both their boundaries due to their total modulating function nature.

%In many parameter estimation and system identification research studies are used concrete predefined modulating functions such as Hartley, splines, trigonometric functions, etc. \cite{jouffroy2015finite}. In contrast to that, our approach is to have the modulating functions defined by the data itself, through the problem formulation. A notable advantage in having this is that the method will define itself, for each parameter, the best match for the modulating function. Very close to this is the work presented in \cite{schmid2011parameteridentifikation} where the parameters of the system are estimated independently.\\

Let us then transform this set of integro-differential equations into integral equations by considering the $n$-th order derivative of $\varphi(t)$ arising in expression (\ref{eq:z}), and define the new function $\alpha : [0,T]\rightarrow \mathbb{R}$ as
 \begin{equation}
\alpha (t) :=  \varphi ^{(n)} (t).
\end{equation}
Then, using the anti-derivative notation
\begin{equation}
	f^{(-i)}(\tau) := \int_{0}^{\tau} \int_{0}^{\tau_i}... \int_{0}^{\tau_2} f (\tau_1) d\tau_1 \, d\tau_2 \,\hdots\, d\tau_i,
\end{equation}
the right boundaries of the derivatives of a total modulating function $\varphi(t)$ can simply be re-written as
\begin{equation}
 \alpha^{(-i)} (T) = 0  \hspace{5mm}, i=\overline{1,n}.
\label{newBC}
\end{equation}
while the zero left boundary conditions are simply fulfilled thanks to the successive integration of $\alpha$.
Thus, integral operator can now be re-written as
\begin{equation}
L^i[y] := \int_{t-T}^{t} (-1)^i \alpha^{(i-n)}(\tau - t + T) y(\tau) d\tau,
\label{Lintegraloperatoralpha}
\end{equation}
 so that $\mathbf{w}$ in (\ref{eq:wi}) only consists of integrals of $\alpha$. Hence, expression (\ref{identityconstraint}) is now a set of integral equations. Grouping the right boundary conditions (\ref{newBC}) similarly, we also get
 \begin{equation}
 \boldsymbol{\Gamma} = \mathbf 0_{n \times m_t},
 \label{boundaryconstraint}
 \end{equation}
 where $ \boldsymbol{\Gamma} = [\boldsymbol{\upalpha}_1,\boldsymbol{\upalpha}_2,...,\boldsymbol{\upalpha}_{m_t}]$, with each vector $\boldsymbol{\upalpha}_j$ composed by all boundary conditions (\ref{newBC}) for this specific modulating function, i.e. $\boldsymbol{\upalpha}_j = [ \alpha_j^{(-n)}, \alpha_j^{(-n+1)}, ..., \alpha_j^{(-1)}]^T$. \\
Combining finally (\ref{identityconstraint}) with (\ref{boundaryconstraint}), we have a system of linear integral equations, which is simple to solve numerically (details of this numerical method are given in Appendix A). Once a function $\alpha(\cdot)$ is obtained for each instant $t$, constraint (\ref{identityconstraint}) is fulfilled, and Gramian-like expression (\ref{Gramianest}) is simply replaced with
\begin{equation}
\hat {\boldsymbol{\uptheta}}(t) =  \frac{1}{T'} \int_{t-T'}^{t}  {\mathbf{z}}(\tau) d\tau.
\label{eq:averagedestimate}
\end{equation}
As for state estimation, it is also possible to normalize $\mathbf{W}_l$ similarly so that we have the state estimate
\begin{equation}
\hat{\mathbf{x}} = \mathbb{O}^{-1} \mathbf z_l.
\end{equation}
The most notable difference with parameter estimation, is that, since $\mathbf{W}_l$ in (\ref{xequationmodified2}) does not depend directly on the measured signals, the $m_l=n$ left modulating functions $\varphi_{l,j}(\cdot)$ can be chosen once and for all $t$.
Note that a similar kind of normalization was also mentioned in the context of least-squares observers (see \cite{539439}).

\subsection{Direct continuous-time implementation}

%\begin{equation}
%\mathbf w =
%\begin{bmatrix} 
%\int_{t-T}^{t} \alpha^{(-1)} (\tau)   y(\tau)\, d\tau  \\
%\int_{t-T}^{t} \alpha^{(-2)} (\tau)   y(\tau)\, d\tau  \\
%\int_{t-T}^{t} \alpha^{(-1)} (\tau)   u(\tau)\, d\tau  \\
%\int_{t-T}^{t} \alpha^{(-2)} (\tau)  u(\tau)\, d\tau  \\
%\int_{t-T}^{t} \alpha^{(-2)} (\tau) \, d\tau  \\
% \end{bmatrix}\text{.}
% \label{wROM}
%\end{equation} 
 
When needing to choose a particular estimation method, one can obviously focus on performance in terms of the best possible match between the output of the considered plant and its corresponding predicted output using the parameter estimates. However, other factors can also be under consideration, such as ease of implementation, portability of the code, etc. Regarding the on-line use of modulating functions, the latter are usually first discretized (as they would be in the previous subsection or as in \cite{co1997batch} and \cite{noack2018road}). However, modern tools for simulation and control such as Matlab/Simulink allows one to program systems and algorithms directly in a continuous-time setting, thus allowing for the additional advantage of being able to consider non-regular samplings. In this section, we present one way to do that, with in mind direct continuous-time implementation, as opposed to looking primarly at performance.\\
To do so, we first begin by introducing a function $\psi :  [0,T]\rightarrow \mathbb{R}$, which we will refer to as \emph{reversed modulating function}, and where $\psi(\cdot)$ is such that 
\begin{equation}
	\psi(t) := \varphi(T-t)
	\label{reversedMF}
\end{equation}
where $\varphi(\cdot)$ is our usual modulating function of the very basic following definition \ref{MFdefinition}. In this case, note that, because of reversal (\ref{reversedMF}), a left reversed modulating function corresponds to a right modulating function, and vice-versa. Then, replace operator (\ref{Lintegraloperator}) with
\begin{equation}
M^i[y] := \int_{t-T}^{t} \psi^{(i)}(t-\tau) y(\tau) d\tau.
\label{Mintegraloperator}
\end{equation}
The advantage of (\ref{Mintegraloperator}), is that, besides its slightly simpler expression than (\ref{Lintegraloperator}), it is directly put under a usual convolution form. Note, also similarly to (\ref{Loperatorproperty}), we have the property that $M^0[y^{(i)}]=M^i[y]$.\\
Next, we notice that many modulating functions defined in the literature can be expressed by the solution of a differential equation. For example, the total modulating function $\varphi(t)=t^2 (t-T)^2$ used for example (\ref{sinewave}) is the solution of differential equation $\varphi^{(4)}(t)=0$. Hence, we introduce the following state-space representation
\begin{align}
\dot {\boldsymbol{\chi}} & = \mathbf{\Lambda}\boldsymbol{\chi}, \, \, \, \boldsymbol\chi(0)=\mathbf{l} \label{chieq}\\ 
\psi & =\mathbf{\Sigma}\boldsymbol{\chi}
\end{align}
where the state vector $\boldsymbol{\chi}\in \mathbb{R}^{n_{\psi}}$.% and where, for simplicity, we assume that 
%\begin{equation}\mathbf{\Lambda}=
%	\begin{bmatrix}
%	0 & 1 & 0 & 0 &\cdots & 0\\ 
%	0 & 0 & 1 & 0 &\cdots & 0\\ 
%	0 & 0 & 0 & 1 &\cdots & 0\\ 
%	\vdots & \vdots & \vdots & \vdots & \ddots & \vdots \\ 
%	0 & 0 & 0 & 0 & \cdots & 1\\ 
%	-p_0 & -p_1 & -p_2 & -p_3 & \cdots  &  -p_{n_{\psi}-1}
%	\end{bmatrix}
%\end{equation}
%and
%\begin{equation}
%\mathbf{\Sigma} =\begin{bmatrix}
%1 & 0 & \cdots & 0
%\end{bmatrix}.
%\end{equation}
Thus, integral operator (\ref{Mintegraloperator}) can easily be rewritten using some advantages of state-space representations. For example, $M^0[y]$ can be rewritten as
\begin{equation}
M^0[y] = \int_{t-T}^{t} \mathbf{\Sigma} e^{\mathbf{\Lambda}(t-\tau)} \mathbf{l} \, y(\tau) d\tau.
\label{Mintegraloperator2}
\end{equation}
and similarly for the other $M^i[y]$'s. Defining now the new vector $\boldsymbol\xi_y \in \mathbb{R}^{n_{\psi}}$ as
\begin{equation}
\boldsymbol\xi_y := \int_{0}^{t} e^{\mathbf{\Lambda}(t-\tau)} \mathbf{l} \, y(\tau) d\tau,
\label{xidef}
\end{equation}
then it can be shown that 
\begin{equation}
	M^0[y] = \mathbf{\Sigma} \left[\boldsymbol{\xi}_y(t)-e^{\mathbf \Lambda T} \boldsymbol{\xi}_y(t-T)\right]
	\label{filterout}
\end{equation}
where vector $\boldsymbol{\xi}_y(t)$ is obtained as the solution of differential equation
\begin{equation}
	\dot{\boldsymbol{\xi}}_y = \mathbf{\Lambda} \boldsymbol{\xi}_y + \mathbf{l} y.
	\label{filterdyn}
\end{equation}
Interestingly, assuming now that system (\ref{chieq}) is stable means that filter (\ref{filterdyn})-(\ref{filterout}) can be used to implement $M^0[y]$ in software tools such as Matlab/Simulink without having to proceed to a preliminary discretization (the same is of course valid for $M^0[u]$, with a state vector $\boldsymbol{\xi}_u$). Note that computing $M^i[y]$ is not more difficult, and we would simply have
\begin{equation}
M^i[y] = \mathbf{\Sigma} \mathbf{\Lambda}^i\left[\boldsymbol{\xi}_y(t)-e^{\mathbf \Lambda T} \boldsymbol{\xi}_y(t-T)\right].
\label{filterouti}
\end{equation}
Gathering now the $M^i[y]$ terms similarly to (\ref{L_lvector}), we have
\begin{equation}
	\mathbf{M}[y] = \mathbb{O}_{MF} \left[ \boldsymbol \xi_y(t) - e^{\mathbf{\Lambda}T} \boldsymbol{\xi}_y(t-T)\right],
	\label{Myfilterout}
\end{equation}
where 
\begin{equation}
\mathbf M[y] = \left[ M^0[y],M^1[y],...,M^{n-1}[y] \right]^T
\label{Mvector}
\end{equation}
and
\begin{equation}
\mathbb O_{MF} = \left[ \mathbf{\Sigma}^T,(\mathbf{\Sigma \Lambda} )^T,...,(\mathbf{\Sigma \Lambda}^{n-1})^T \right]^T.
\label{ObsMF}
\end{equation}
From there, we end up with regression (\ref{eq:systemOfE}) again, where
\begin{equation}
	\mathbf{w} = \left[\mathbf{M}^T[y],\mathbf{M}^T[u]\right]^T
	\label{newregression}
\end{equation}
while $z$ is given by
\begin{equation}
	z=M^n[y].
\end{equation}
In case one wants to favor speed over precision, it is possible to use a single total (reversed) modulating function and use a Gramian-based expression in order to get the parameter estimate. Another advantage of using expressions such as (\ref{Mintegraloperator2}) or (\ref{filterout}), where the modulating function is generated by a stable system, is that similar expressions can also be used to obtain the parameter estimates themselves. Indeed, as in \cite{5400719}, we can use a generalized Gramian expression where the kernel is generated thanks to a reversed modulating function. Indeed, multiply each term of (\ref{eq:systemOfE}) by $\psi(t-\tau)\mathbf{w}(\tau)$ and integrate to get
\begin{equation}
\int_{t-T}^{t} \psi(t-\tau) \mathbf{w}(\tau)z(\tau) d\tau = \int_{t-T}^{t} \psi(t-\tau) \mathbf{w}(\tau)\mathbf{w}^T(\tau) d\tau \, \boldsymbol{\uptheta}
\label{Mintegralequality}
\end{equation}
which can be rewritten as
\begin{equation}
	M^0[\mathbf{h}] = M^0[\mathbf{G}] \boldsymbol{\uptheta}
\end{equation}
where we have vector $\mathbf{h}=\mathbf{w}z$ and matrix $\mathbf{G}=\mathbf{w}\mathbf{w}^T$. Proceeding in the same manner as we did from expressions (\ref{Mintegraloperator}) to (\ref{Myfilterout}), we can define, for $M^0[\mathbf{h}]$, the filter equations
\begin{align}
\dot {\boldsymbol{\xi}}_{\mathbf{h}} & = \underline{\mathbf{\Lambda}}\boldsymbol{\xi}_{\mathbf{h}} + \underline{\mathbf{l}} \mathbf{h} \label{chiheq}\\ 
M^0[\mathbf{h}] &  = \underline{\mathbf{\Sigma}} \left[\boldsymbol{\xi}_{\mathbf{h}}(t)-e^{\underline{\mathbf\Lambda} T} \boldsymbol{\xi}_{\mathbf{h}}(t-T)\right]
\end{align}
where $\boldsymbol{\xi}_{\mathbf{h}} \in \mathbb{R}^{2nn_{\psi}}$, and where $\underline{\mathbf{\Lambda}} = \mathbf\Lambda \otimes \mathbf I_{2n}$, 
$\underline{\mathbf{l}} = \mathbf l \otimes \mathbf I_{2n}$ and $\underline{\mathbf{\Sigma}} = \mathbf\Sigma \otimes \mathbf I_{2n}$ (with $\otimes$ for the Kronecker symbol). For $M^0[\mathbf{G}]$, we have same kind of filter, this time with a matrix differential equation
\begin{align}
\dot {\boldsymbol{\Xi}}_{\mathbf{G}} & = \underline{\mathbf{\Lambda}}\boldsymbol{\Xi}_{\mathbf{G}} + \underline{\mathbf{l}} \mathbf{G} \label{Chieq}\\ 
M^0[\mathbf{G}] &  = \underline{\mathbf{\Sigma}} \left[\boldsymbol{\Xi}_{\mathbf{G}}(t)-e^{\underline{\mathbf\Lambda} T} \boldsymbol{\Xi}_{\mathbf{G}}(t-T)\right]
\end{align}
where $\boldsymbol{\Xi}_{\mathbf{G}} \in \mathbb{R}^{ 2nn_{\psi} \times 2n}$. Finally, an estimate of parameter vector $\boldsymbol{\uptheta}$ is obtained by simple inversion as
\begin{equation}
	\hat{\boldsymbol{\uptheta}} = \left( M^0[\mathbf G] \right)^{-1} M^0[\mathbf h].
	\label{directparameterestimation}
\end{equation}

 Interestingly, and moving now to state estimation, the method simply consists in repeating the steps (\ref{xequationmodified}) to (\ref{xhateq}) by taking into account that left modulating integral operators $L^i_l[y]$ and $L^i_l[u]$ are replaced by the right reversed modulating integral operators $M^i_r[y]$ and $M^i_r[u]$, where we have the property
 \begin{equation}
 M^0_r[y^{(i)}] = \sum_{k=0}^{i-1} \psi_r^{(k)}(0)y^{(i-1-k)}(t) + M^i_r[y]
 \end{equation}
(note, therefore, that steps (\ref{newregression}) through (\ref{directparameterestimation}) are obviously not necessary).

\section{Case study}
\label{section:casestudy}

As a case study, a heat flow experimental chamber produced by Quanser company is considered. This plant reproduces the thermodynamics of an Air Handling Unit of an HVAC system. A few technical details related to the experimental set-up will be first introduced, while results of on-line parameter and state estimation and their validation will be presented afterwards. 

\subsection{Quanser heat flow chamber}

\begin{figure}[!htb]
\centerline{
    \includegraphics[width=8.5cm]{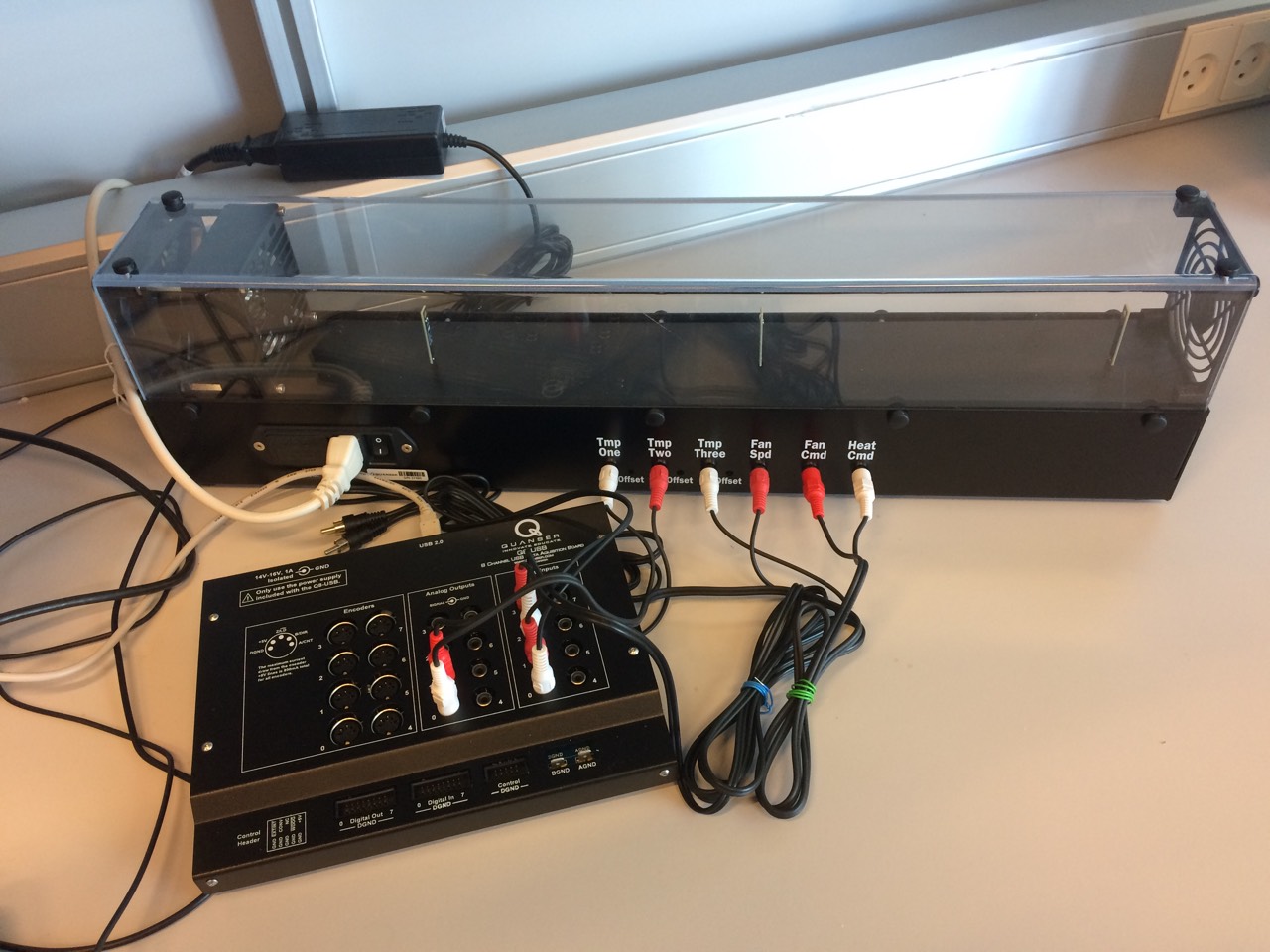}
    }
\caption{Quanser Heat Flow Experimental setup}
\label{HFE}
\end{figure}

The heat flow experiment (Fig. \ref{HFE}) consists of a fiber-glass chamber equipped with a fan blowing over an electric heating coil. Both the fan and the heater can be controlled externally with a $0-5 V$ input signal. The air temperature inside the box is measured by three temperature sensors positioned equidistantly. Additionally, we make use of the Quanser Q8-USB Data Acquisition Board in order to enable the communication between the computer and the experimental chamber.

%The Quanser heat flow experiment is mostly used for controller design applications. Usually, the model is defined using a first-order transfer function \cite{MALEK2013746}, \cite{yamada2005modified}. %Another interesting work based on web accessibility was presented in \cite{VARGAS2006141}. The specific application developed for automation technicians training offered remote access via Internet for the actuators and the sensors. \\
%
%{\color{blue} Not sure if we need the above paragraph.}

\subsection{Initial set-up and simulation environment}

In the literature, the considered system model for control applications is usually of first order model (with or without time delay) \cite{MALEK2013746}\cite{yamada2005modified}. However, in practice, a second-order model can have its importance, as the surface does not only represent the envelope/box storing heat but also other components (fans, dampers, filters). In a faulty situation where the box /components inside get overheated and can be damaged, estimating the box temperature can generate important information, and help in fault detection.
%For this purpose, Our approach towards modeling is to use a second order ROM. For this specific experiment, we want to directly estimate the parameters of the model first, and perform state estimation afterwards. 
The set-up will be used both for parameter estimation and state estimation scenarios where the latter uses the results of the former.
Because the chamber is located in a closed indoor environment and the experiment is conducted on short time interval, we will assume that the room temperature  ($T_r$) is unknown but constant during the experiment. In order to stay close to an actual operation of a constant flow AHU, the fan is operating with a constant speed during the whole experiment.

\subsection{On-line parameter estimation}

%In a previous study, for the same heat flow model, we have assumed the room temperature $T_r$ in eq. (\ref{eq:2}) as a known input. See \cite{AIM_ANA_Jerome} for more details. In contrast to that, now we treat $T_r$ as an unknown constant disturbance. 

Using steps (\ref{eq:compactE}) to (\ref{eq:system}) on state-space representation (\ref{eq:3})-(\ref{Cmatrix}), we obtain the ordinary differential equation
\begin{equation}
y^{(2)} =  - a_1 y^{(1)}- a_0 y + b_{1} u^{(1)}  + b_{0} u + d
\label{eq:ODE_syst}
\end{equation}
where $y(t)=T_m(t)$ is the measured output, $u(t)$ is the input $Q_{h}$, and $a_1$, $a_0$, $b_1$, $b_0$ and $d$ are the unknown coefficients of the equation, the last one, $d$ representing the impact of the unknown input $T_r$ mentioned above, and considered here as a disturbance. These coefficients are related to the parameters of system (\ref{eq:1})-(\ref{eq:2}) as follows:
\begin{equation}
a_0 = \frac{1}{C_{m}R_{ms}C_{s}R_{sr}} 
\label{eq:a0}
\end{equation}

\begin{equation}
a_1 = \frac{1}{C_{s}R_{ms}} + \frac{1}{C_{s}R_{sr}} + \frac{1}{C_{m}R_{ms}} 
\label{eq:a1}
\end{equation}

\begin{equation}
b_{0} = \frac{1}{C_{m}C_{s}R_{ms}} + \frac{1}{C_{m}C_{s}R_{sr}}
\label{eq:b10}
\end{equation}

\begin{equation}
b_{1} = \frac{1}{C_{m}}
\label{eq:b11}
\end{equation}

\begin{equation}
d = \frac{1}{C_{m}R_{ms}C_{s}R_{sr}}T_r
\label{eq:b20}
\end{equation}

To perform the parameter estimation we have used the integration period $T = 2000\ sec $ with a sampling time $T_s = 2\ sec $, thus giving a total of $1000$ samples. The additional receding horizon interval for obtaining the estimates is $T' = 2000\ sec $.

\subsubsection*{Input profile and persistence of excitation}

A key point in parameter estimation is the persistence of excitation of the input signal. Reliable estimates will be obtained if the input signal is sufficiently rich so that the observed response contains the required information to perform the estimation process. As specified in \cite{Ljung:1999:SIT:293154}, if a specific matrix characteristic of the input signal is non-singular, the input is considered to be persistent.  
%Some very common definitions of persistent inputs for system identification are multi-sine signals, Pseudo-Random Binary signal (PRBS) or Pseudo-Random Sequences \cite{braun2002application}. 
It is well-known that, when estimating the parameters of a system with an input signal having enough persistence of excitation, the estimated parameters approach their true values \cite{481517}. However, it is worth mentioning that, unfortunately, not enough attention has been given to the consideration of persistent inputs in building models and thus not many studies can be found on this topic \cite{li2014review}.

\begin{figure}[!htb]
\centerline{
    \includegraphics[trim=0cm 0cm  0cm 0cm, clip, width = \textwidth]{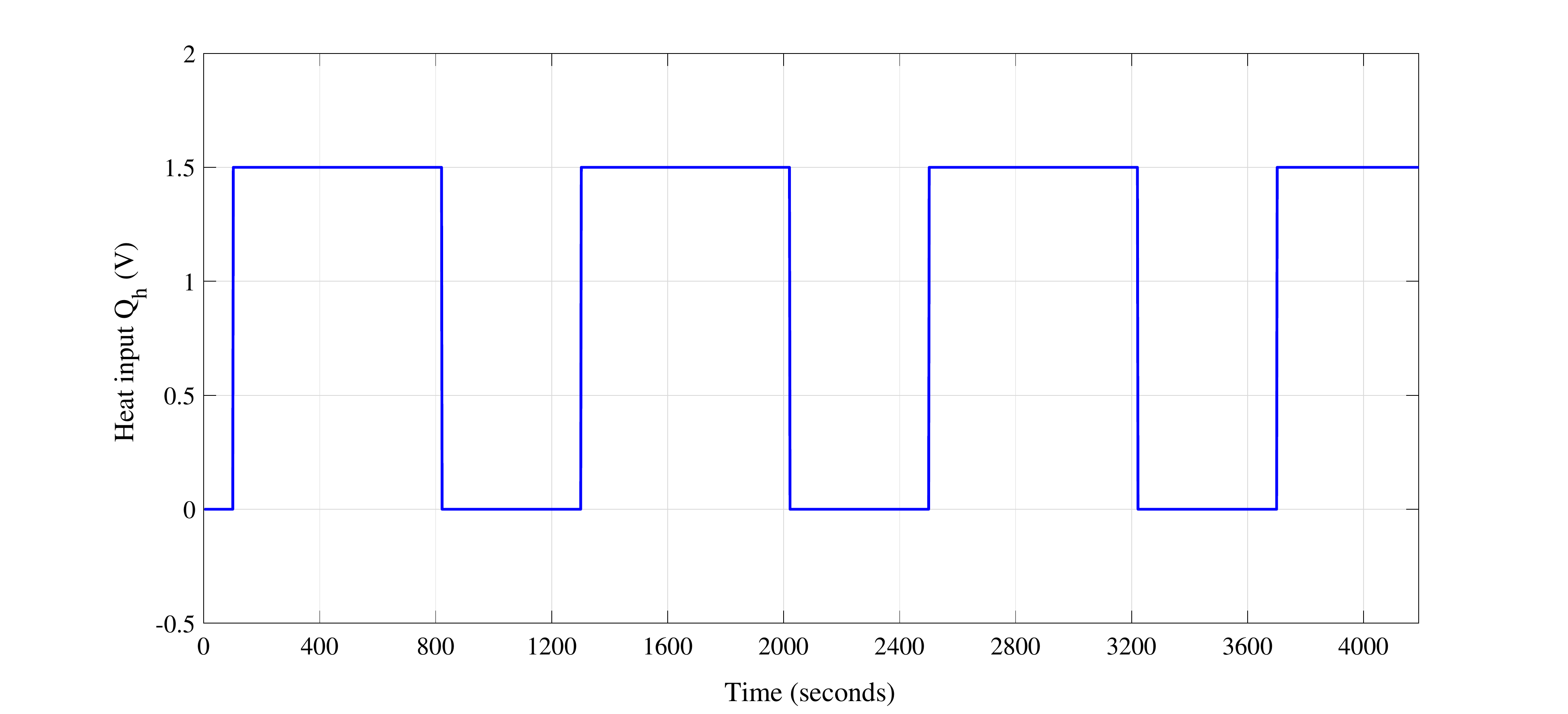}
    }
\caption{Input profile for parameter estimation}
\label{PE_input}
\end{figure}

The considered input signal for parameter estimation is a pulse function shown in Fig. \ref{PE_input}. This signal is informative enough to capture the dynamics of the heating chamber and obtain the estimates. The amplitude of signal is selected to $1.5 V$ to change the temperature of chamber up to around $30^{\circ C}$. The period of the signal is $1200\ sec$ which is long enough for the plant to reach at least $80 \%$ of its final value.\\ 
%(which is basically a first order PRBS). 

\subsubsection*{Estimates obtained using the modulating function}

\begin{figure*}[!htb]
\centerline{
    \includegraphics[trim=0cm 0cm  0cm 0cm, clip,  width=1\textwidth]{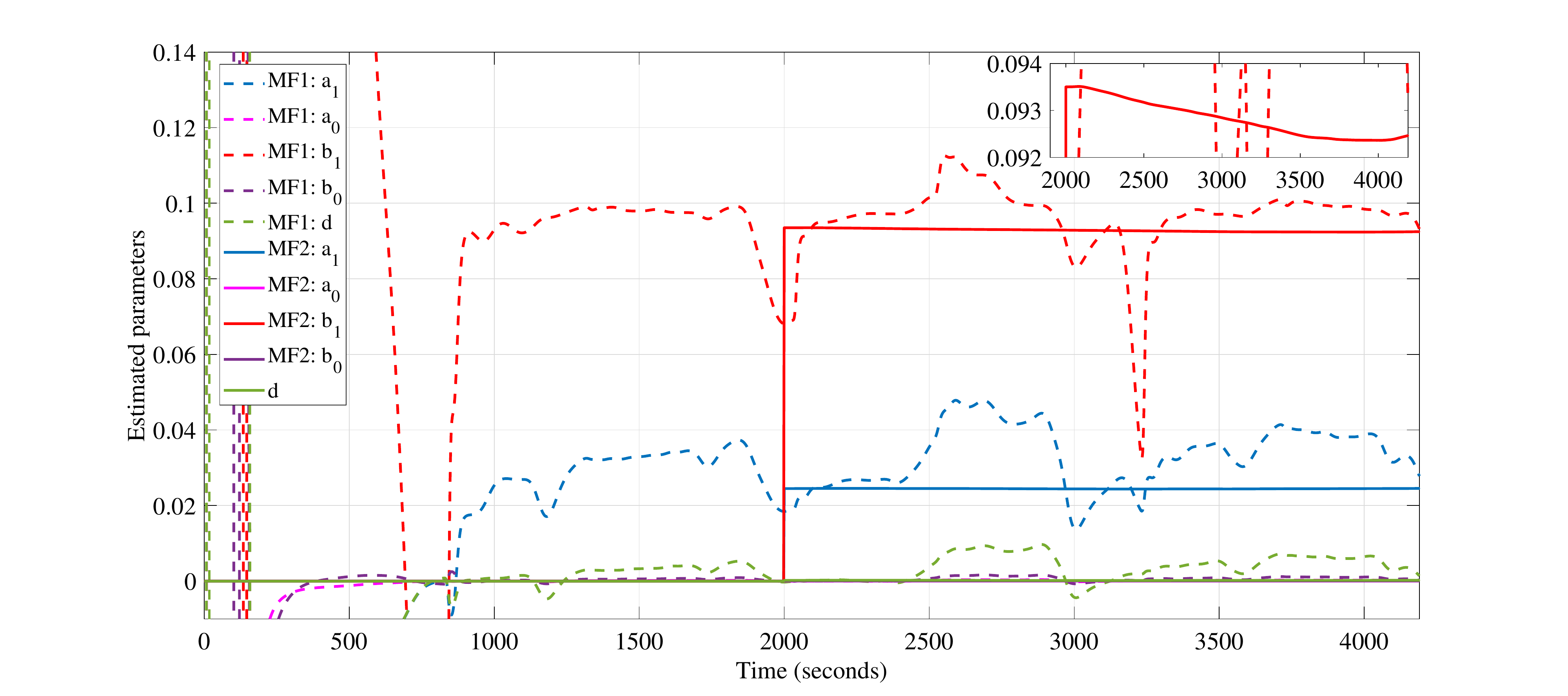}
    }
\caption{Estimated parameters}
\label{PE_parameters}
\end{figure*}

The results of parameter estimation using the modulating function method described in section \ref{section:MF} are presented in Fig. \ref{PE_parameters}.
The results are given for both methods: the time-vayring modulating function technique (continuous lines) and the direct continuous-time technique (dashed lines). As expected, the direct continuous-time technique, using only one modulating function, does not perform as well as the time-varying technique (although it is computationally more efficient). The latter shows quite good results and converges to an almost constant value after the additional receding horizon interval $T'$.

%The results for the estimated parameters using the method described in section \ref{section:MF} are presented in Fig. \ref{PE_parameters}. 
%We can observe that when using several modulating functions (parameters marked with dashed lines) we obtain the parameters which have almost a fixed value after the first additional receding horizon interval $T'$. In contrast to this, when using only one modulating function (parameters marked with continuous line), even if the convergence is comparable good, the fact that we don't have a unique modulating function for each parameter creates undesirable fluctuations. 
\begin{figure*}[!htb]
\centerline{
    \includegraphics[trim=1cm 0cm  0cm 0cm, clip,width=01\textwidth]{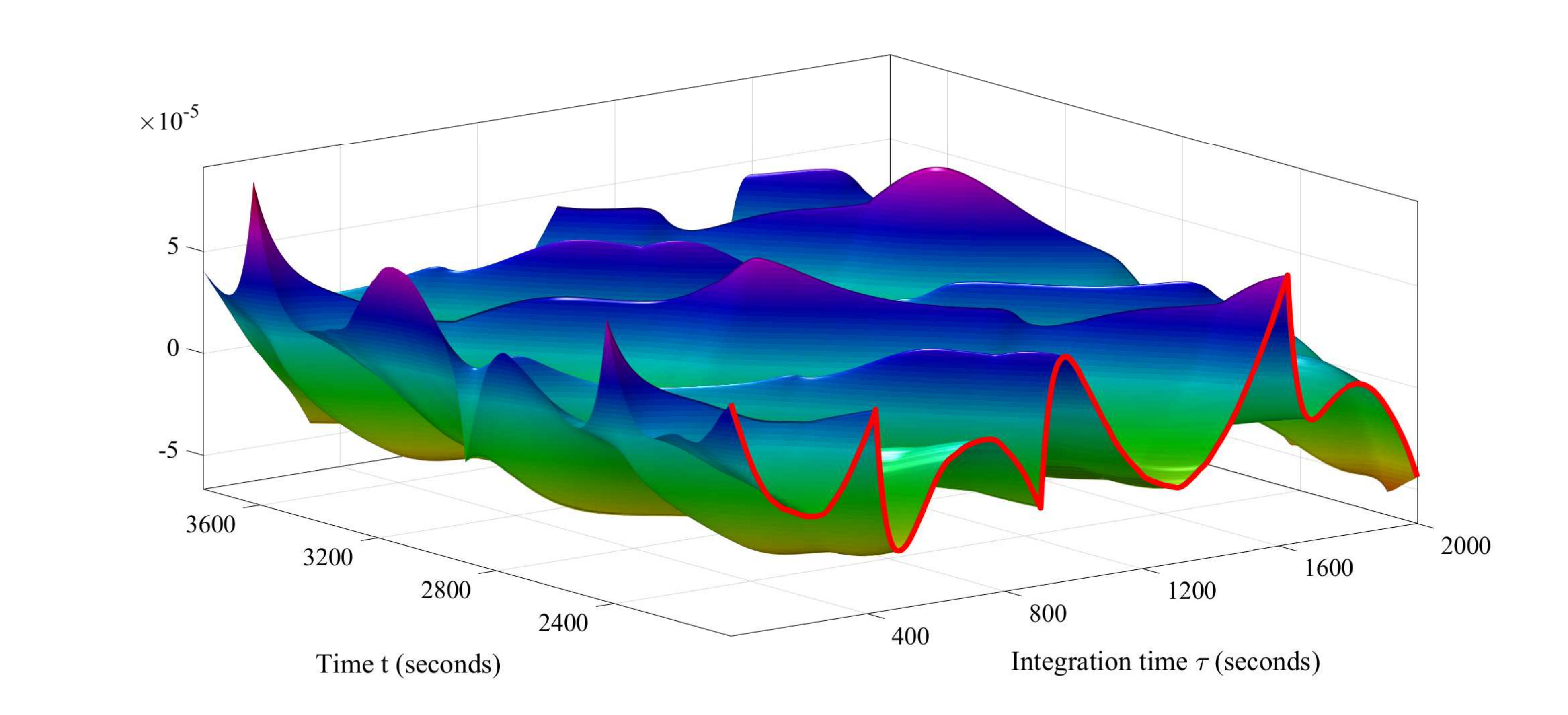}
    }
\caption{Evolution of the modulation function}
\label{MF_surf}
\end{figure*}
%An overview of how the modulating function evolves in time is shown in Fig. \ref{MF_surf}. 

Relating to the time-vayring modulating function method presented in section \ref{MFcalculation}, Fig. \ref{MF_surf} shows how one of the total modulating functions (corresponding in this case to $a_1$) evolves over time. The modulating function is a signal of length $T=2000 sec$, and the figure shows its evolution from $t=2000$ to $t=3700$ (the curve highlighted in red is MF at time obtained $t = 2000 sec$). As it can be seen, the amplitude of the MF is constantly adapting to new incoming data/measurements that are fed on-line to the estimator.
 
\subsection{Validation}
Since there is no upfront knowledge about the real values of the parameters, we will compare the results of the modulating function method with the results of two well-known estimation methods available in the System Identification Toolbox for Matlab \cite{ljung2007system}. These two methods, applicable to continuous-time systems, are the Transfer Function method and the Grey-box estimation method. Moreover, for further validation of the results, the system output is reconstructed using the estimated parameters compared with the real measurements.\\

\subsubsection{Continuous Transfer Function Estimation}
\label{sec:TF}
The first method used for comparison is the ``tfest'' function of Matlab which estimates a continuous transfer function for the given data. In this method the input and output of the model are filtered using a pre-filter $L(s)$, and the differential equation of the system is written in a regression form with respect to the filtered signals. The parameters are then estimated to minimize the prediction error. \par
Matlab uses different numerical search methods to estimate the parameters iteratively. In case of selecting the default setting, a combination of four line search algorithms  'Subspace Gauss-Newton least squares search', 'Levenberg-Marquardt least squares search', 'Adaptive subspace Gauss-Newton search', and 'Steepest descent least squares search' methods will be applied and the first descent direction leading to a reduction in estimation cost is used.\par
In order to initialize the parameters, different algorithms are available in Matlab. The algorithm used in this study is the instrument variable approach in which the parameters are estimated using the least-squares method \cite{garnier2003continuous} and \cite{ljung2009experiments}.
\subsubsection{Grey-Box Model Estimation}
The second method used for comparison is the grey-box model estimation using the ``idgrey'' function of Matlab. This method directly uses the parametric state-space representation of the system given in (\ref{eq:3})-(\ref{eq:measure}) and minimizes the prediction error.\\
Matlab uses the same numerical search methods as explained in section \ref{sec:TF}. However, contrary to the transfer function estimation technique, the initial value for the parameters should be given by the user.\par

\begin{table}[htb!]
  \centering
\caption{Summary of estimated parameters}
\hspace{0.5cm}  \begin{tabular}{ |l | p{1.7cm} |  p{1.7cm}|  p{1.7cm}|  }
    \hline
&Grey-box&Transfer \hspace{0.5cm}Function&Modulating Function\\ \hline
 \textbf{$a_0$}&1.2312e-06&1.454e-05&9.958e-6\\ \hline
 \textbf{$a_1$}&0.0258&0.02447&0.02449\\ \hline
\textbf{$b_0$} &4.1891e-05&7.385e-05&6.81e-5\\ \hline
 \textbf{$b_1$} &0.098591&0.09188&0.09236\\ \hline
\textbf{$d$}&1.2312e-06 &3.3891e-04&2.1770e-04\\ 
\hline
  \end{tabular}
\label{tab:Parameters}
\end{table}

\begin{figure*}[!htb]
\centerline{
   \hspace*{0.5cm} \includegraphics[width=1\textwidth]{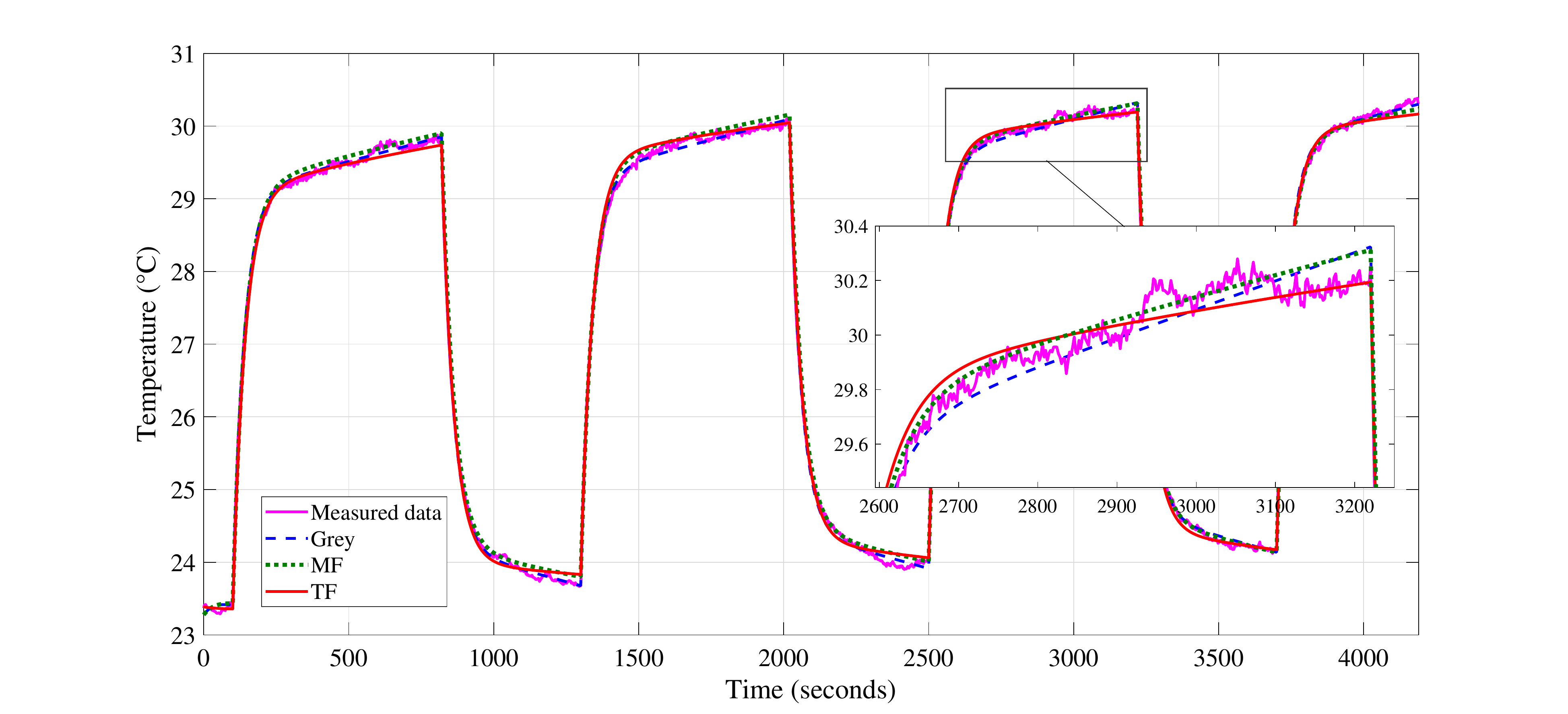}
    }
\caption{Validation - measured vs. simulated model output}
\label{results:comparison}
\end{figure*}

Table \ref{tab:Parameters} shows the estimated parameters using the MF method together with those obtained by the transfer function and grey-box approaches. As it can be seen, the results of the MF method are very close to the results of the transfer function method, while there is a discrepancy in some parameters with when compared to the grey-box method. The comparison is more clear in Fig. \ref{results:comparison} where the output of the estimated models are plotted together and compared with the measured  data. The goodness of the fit is $96.95 \%$, $96.51 \%$ and $97.42 \%$, for the modulating function, transfer function and gray-box methods, respectively, which confirms the validity of the proposed MF method. \par
%The estimated parameters for the considered 2nd order model are summarized in Table \ref{tab:Parameters}. 
The main reason for achieving a slightly different (better) result by grey-box estimation method is that the reported goodness of fit was based on non-filtered measured data which matches the structure of grey-box method. The grey-box method works on the original input-output signals whereas the other two methods, i.e. modulating function and transfer function methods, use a pre-filter for the input-output signals ($L(s)$ in transfer function method and the modulating integral operators in the MF method). Therefore the optimization algorithm in the MF and transfer function methods minimizes the filtered error while the reported goodness of fit was based on the original error. In case we would define the goodness of fit based on filtered signals, a higher fit would be achieved for MF and transfer function methods. \par
Another important point that should be mentioned is that our implementation of the MF method is on-line while the other two methods are iterative and off-line. In particular, the convergence of the grey-box method highly depends on the initial guess for the parameters (in this study, the estimated parameters from MF method are considered as initial guess for the grey-box method). Therefore, achieving a result in the order of off-line methods in finite time further attests the acceptable performance of our results.

%The comparison of the step and frequency response of the three estimated methods are given in Figure(8). As it is observed, the main discrepancy of the results is in the low frequencies (steady state value).  represent the 

%\textcolor{red}{Hossein please explain why the gray-box is the best and how was obtained}

%\begin{figure}[htb!]
%\centering
%\subfigure{
%	\includegraphics[width=0.5\textwidth]{Fit_TF} }  
%\subfigure{
%	\includegraphics[width=0.5\textwidth]{Fit_MF} } 
%\subfigure{
%	\includegraphics[width=0.5\textwidth]{Fit_GB}} 
%\caption{Model fit}
%\label{Fit}
%\end{figure}

%\begin{figure}[!htb]
%\centerline{
%   \hspace*{0.5cm}\includegraphics[width=0.5\textwidth]{Compare_bode}
%    }
%\caption{}
%\label{}
%\end{figure}

%\begin{figure}[!htb]
%\centerline{
% \includegraphics[width=0.5\textwidth]{Compare_PE_stepResponse_newL}
%    }
%\caption{Step response}
%\label{}
%\end{figure}

%\textcolor{red}{Hossein please explain Step response figure\\}

\subsection{On-line state estimation}

Given the estimated parameters from previous section, it is straightforward to apply the method described in section \ref{section:MF} to obtain the state estimates. Here, both $T_m$ and $T_s$ are estimated on-line using an integration interval $T = 50 sec$ with sampling period $T_s = 1 sec$.

\begin{figure*}[!htb]
\centerline{
    \includegraphics[trim=0cm 0cm  0cm 0cm, clip,width=1\textwidth]{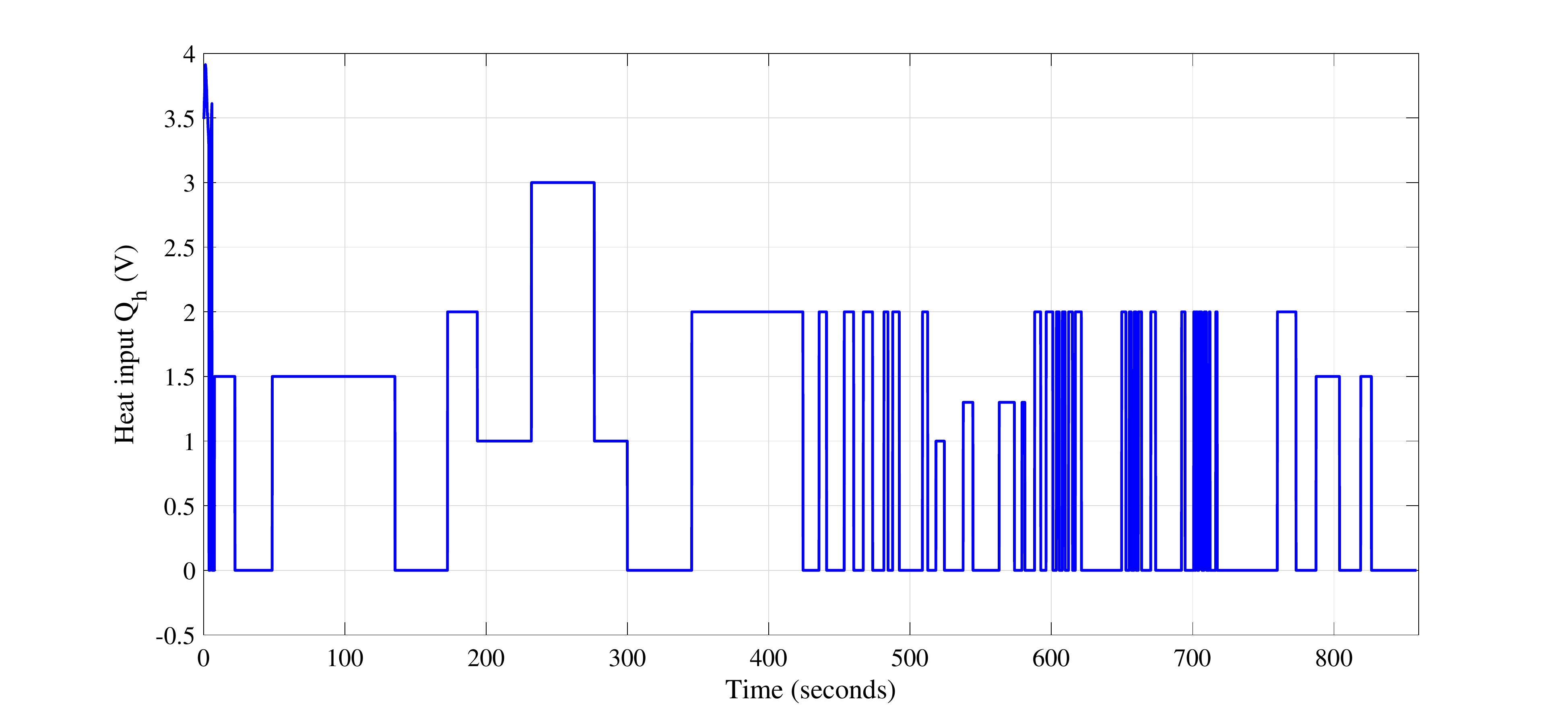}
    }
\caption{Input profile for state estimation}
\label{SE_input}
\end{figure*}

In a normal operation set-up, the requirements for the temperature at the exit of the AHU might not vary too much in a short time interval. However, to test the proposed state estimation algorithm in different work conditions, a pseudo random input profile shown in Fig. \ref{SE_input} is applied to the system. This is not a common operating profile, but it helps to visualize the state estimator and evaluate its behavior. 

\begin{figure*}[!htb]
\centerline{
    \includegraphics[trim=0cm 0cm  0cm 0cm, clip, width=1\textwidth]{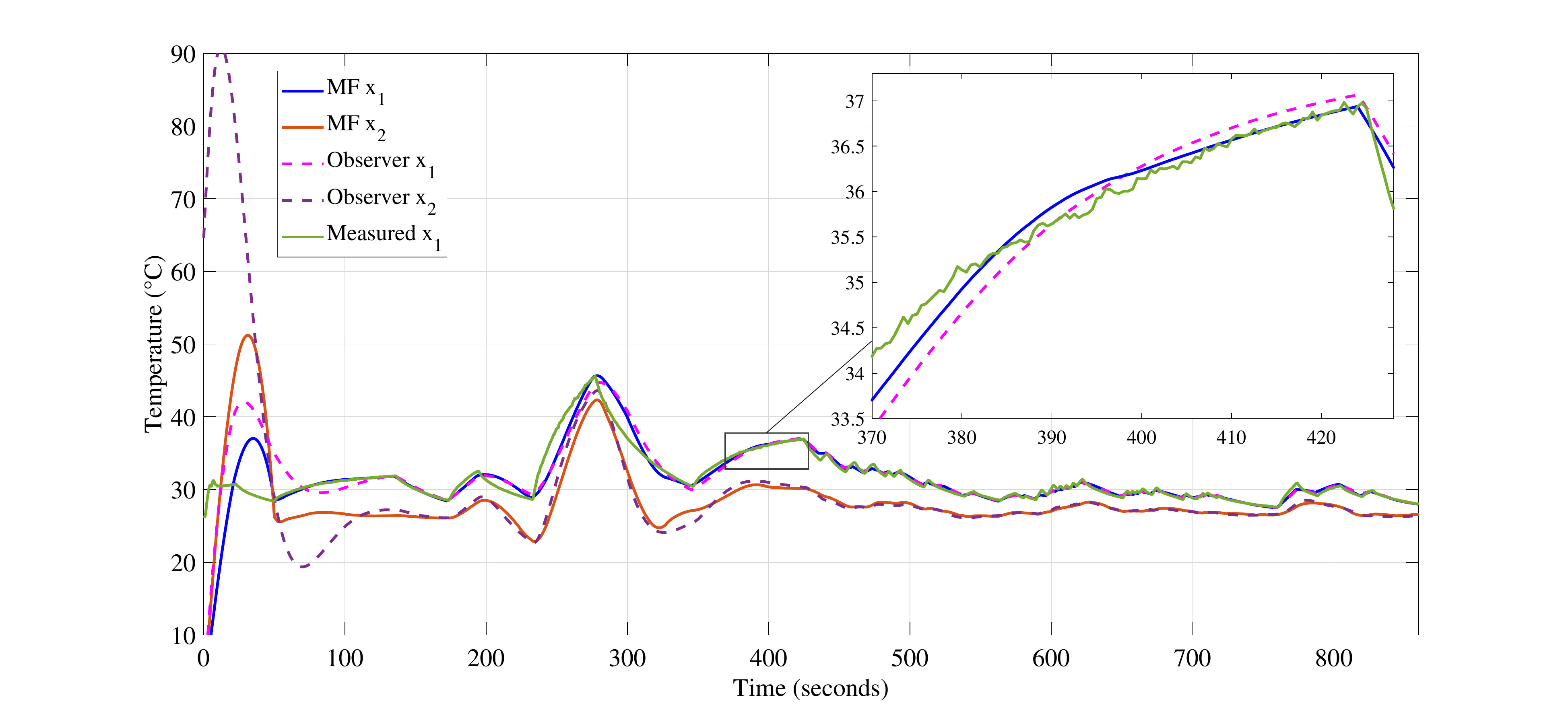}
    }
\caption{Measured vs. estimated temperature inside the heating chamber using the modulating function and simple observer.}
\label{SE_states}
\end{figure*}

In Fig. \ref{SE_states}, it can be observed that the estimated temperature inside the chamber is very close to the noisy measured temperature. Also, as expected, the value for the state representing the envelope temperature of the chamber is lower than the temperature inside, and follows the same heating profile which is in perfect concordance with the heating input profile. The estimated states are comparably good with the ones obtained from a standard observer. In this specific case, the convergence of the states is faster due to the fixed integration interval, which ensures convergence after the first receding horizon window.

\section{Conclusion}
\label{section:conclusions}

A 2nd order model is presented in this paper to represent the heat flow in the air handling unit present in compact HVAC systems. To facilitate the real-time applications of the model we have proposed an unconventional method, the modulating function, with two different implementation approaches.

As a case study, a heat flow experiment by Quanser, whose thermodynamics are very close to those of an AHU, is considered. The on-line parameter and state estimation algorithms were implemented in Matlab/Simulink. The results underline the potential of this approach, showing a good match between the measured and estimated parameters and states. A good convergence of the parameters was observed having as a baseline the parameters estimated with standard estimation methods: the transfer function method and the grey-box method. Moreover, using several (as opposed to only one) modulating functions gives a better estimation of the parameters compared with an approach that uses only one, at the cost of speed although the latter one mentioned requires less resources to implement and run. The method can be successfully used for on-line state estimation as well, the results being comparable good with a standard observer.

% or


\section*{Appendix A - Numerical integration}
\label{Appendix1}

For implementing the proposed estimation method a simple Riemann sum is used, taking into account that the signals $y(t)$ and $\mathbf u(t)$ are in practice sampled at regular intervals. Therefore, for the simple boundary condition (\ref{newBC}), with $i=1$, we use the approximation
\begin{equation}
\alpha^{(-1)}(T) = \int_{t-T}^{t} \alpha (\tau - t +T) \, d\tau   \approx   \displaystyle\sum_{k=1}^{N}T_{s} \underline{\upalpha}(k) = T_{s} {\mathbf 1}^T \underline{\boldsymbol{\upalpha}}
\label{eq:startT}
\end{equation}
where $N$ is the number of samples over the interval, $T_s$ is the sampling period, and $\underline{\upalpha}(k)$ is the sampled value of function $\alpha$ at iteration $k$, which gives the vector $\underline{\boldsymbol{\upalpha}}^T=[\underline{\upalpha}(1),\underline{\upalpha}(2),...,\underline{\upalpha}(N+1)]^T$. Vector ${\mathbf 1}$ is a vector of dimension $N$ containing only ones. Using a similar reasoning, $\alpha^{(-1)}(\tau)$ can be approximated as
\begin{equation}
\alpha^{(-1)}(\tau) = \int_{t-T}^{\tau}\alpha(\sigma - t +T) d\sigma  \approx   \displaystyle\sum_{l=1}^{k}T_{s} \underline{{\upalpha}}(l)=: \underline{{\upalpha}}^{(-1)}(k)
\end{equation}
so that 
\begin{equation}
	\underline{\boldsymbol{\upalpha}}^{(-1)} = T_s \mathbf Q \underline{\boldsymbol{\upalpha}}
\end{equation}
where $\underline{\boldsymbol{\upalpha}}^{(-1)T}=[\underline{{\upalpha}}^{(-1)}(1),\underline{{\upalpha}}^{(-1)}(2),...,\underline{{\upalpha}}^{(-1)}(N + 1)]^T$, while the matrix $\mathbf Q$ is a lower triangular matrix of ones given by
\begin{equation}
\mathbf Q =
\begin{bmatrix} 
1  & 0& \dots& 0\\
\vdots  & \ddots&\ddots& \vdots\\
1 &\dots&  1&0\\
1  & 1&\dots& 1\\
\end{bmatrix}.
\end{equation}
Hence, for $i=2$, condition (\ref{newBC}) gives
\begin{equation}
	\alpha^{(-2)}(T) = \int_{t-T}^{t} \alpha^{(-1)} (\tau -t +T) \, d\tau   \approx T_s^2 {\mathbf 1}^T \mathbf Q \underline{\boldsymbol{\upalpha}}.
\end{equation}
 and for $i=\overline{3, n}$:

\begin{equation}
	\alpha^{(-i)}(T) = \int_{t-T}^{t} \alpha^{(-i + 1)} (\tau -t +T) \, d\tau   \approx T_s^i {\mathbf 1}^T \mathbf Q^{i-1} \underline{\boldsymbol{\upalpha}}.
\end{equation}
The operators $L^i[y]$ of $\mathbf w$ in (\ref{Lintegraloperatoralpha}) can be similarly approximated, so that 

\begin{equation}
\begin{split}
L^i[y] := \int_{t-T}^{t} (-1)^i \alpha^{(i-n)}(\tau - t + T) y(\tau) d\tau \\
\approx (-1)^i T_s^{n-i+1} \underline{{\mathbf y}}^T\mathbf Q^{n-i}\underline{\boldsymbol{\upalpha}},  i=\overline{0, n-1}
\end{split}
\label{App:Li}
\end{equation}

%\begin{equation}
%\int_{t-T}^{t} \alpha^{(-n)} (\tau)   y(\tau)\, d\tau \approx   T_{s}^{n+1}  \underline{{\mathbf y}}^T \mathbf Q^n \underline{\boldsymbol{\upalpha}}
%\label{Ia-1}
%\end{equation}
%
%for $w_1$, where $\underline{{\mathbf y}}$ is the vector of samples of continuous-time signal $y(t)$. The expression for $w_2$ is approximated to
%\begin{equation}
%\int_{t-T}^{t} (-1)\alpha^{(-n+1)} (\tau)   y(\tau)\, d\tau \approx  (-1) T_{s}^n  \underline{{\mathbf y}}^T \mathbf Q^{n-1} \underline{\boldsymbol{\upalpha}}
%\label{Ia-2}
%\end{equation}
%
%and for $w_n$ term
%
%\begin{equation}
%\int_{t-T}^{t} (-1)^{n-1}\alpha^{(-1)} (\tau)   y(\tau)\, d\tau \approx  (-1)^{n-1} T_{s} \underline{{\mathbf y}}^T \mathbf Q \underline{\boldsymbol{\upalpha}}
%\label{Ia-n}
%\end{equation}

Likewise are defined the remaining terms of $\mathbf w$ in (\ref{Lintegraloperatoralpha}), where vector $\underline{{\mathbf y}}$ in (\ref{App:Li}) is replaced by $\underline{{\mathbf u}}$. Taking now $m_t=2n$ modulating functions, $\mathbf W$ in (\ref{identityconstraint}) can thus be approximated by
\begin{equation}
	\mathbf W \approx \underline{\mathbf W} = \underline{\mathbf K} \, \underline{\underline{\upalpha}}
\end{equation}
where $\underline{\underline{\boldsymbol{\upalpha}}}=[\underline{{\boldsymbol{\upalpha}}}_1,\underline{{\boldsymbol{\upalpha}}}_2,...,\underline{{\boldsymbol{\upalpha}}}_{m_t}]$ and the matrix $\underline{\mathbf K}$ is given by
\begin{equation}
\underline{\mathbf K} =
\begin{bmatrix} 
(-1)^{0}T_{s}^{n+1}  \underline{{\mathbf y}}^T \mathbf Q^{n}  \\
\vdots \\
(-1)^{n-1}T_{s}^2  \underline{{\mathbf y}}^T \mathbf Q  \\
(-1)^{0}T_{s}^{n+1}  \underline{{\mathbf u}}^T \mathbf Q^{n}  \\
\vdots \\
(-1)^{n-1}T_{s}^2  \underline{{\mathbf u}}^T \mathbf Q  \\
\end{bmatrix}.
\label{Ku}
\end{equation}		
Proceeding similarly with the discrete approximation of $\mathbf \Gamma$ in (\ref{boundaryconstraint}), it can be expressed
\begin{equation}
	\mathbf \Gamma \approx \underline{\mathbf \Gamma} = \underline{\mathbf B} \, \underline{\underline{\upalpha}}
\end{equation}
where
\begin{equation}
\underline{\mathbf B} =
\begin{bmatrix} 
T_s^n {\mathbf 1}^T \mathbf Q^{n-1}    \\
\vdots \\
T_s^2 {\mathbf 1}^T \mathbf Q    \\
 T_{s} {\mathbf 1}^T 
\end{bmatrix}.
\label{Bu}
\end{equation}
Then, matrix $\underline{\underline{\boldsymbol{\upalpha}}}$ is obtained by simple pseudoinversion, i.e.
\begin{equation}
\underline{\underline{\boldsymbol{\upalpha}}} = 
\begin{bmatrix}
\underline{\mathbf K} \\ 
\underline{\mathbf B}
\end{bmatrix}^+
\begin{bmatrix}
\mathbf I_{m_{\phi}} \\ 
\mathbf 0_{n \times m_{\phi}}
\end{bmatrix}.
\end{equation}
The discrete approximation of $\mathbf z$ is given by
\begin{equation}
	\mathbf z \approx \underline{\mathbf z} = T_s \underline{{\mathbf y}}^T \underline{\underline{\boldsymbol{\upalpha}}} .
\end{equation}

The parameter estimate vector is finally obtained after applying the simple discretized version of receding-horizon expression (\ref{eq:averagedestimate}).

%\section{Choosing the modulating functions for state estimation}
%\label{Appendix2}

%\ifCLASSOPTIONcaptionsoff
%  \newpage
%\fi



%%\bibliographystyle{IEEEtranS}
%\bibliographystyle{apa}
%\bibliographystyle{IEEEtr}
\bibliography{ms}

\end{document}